# Retrieval of non-sparse object through scattering media beyond the memory effect

Meiling Zhou[1], An Pan[1], Runze Li[1], Yansheng Liang[2], Junwei Min[1], Tong Peng[1], and Chen Bai[1], Baoli Yao[1]*

[1] State Key Laboratory of Transient Optics and Photonics, Xi'an Institute of Optics and Precision Mechanics, Chinese Academy of Sciences, No.17 Xinxi Road, Xi'an 710119, China

[2] Shaanxi Key Laboratory of Quantum Information and Quantum Optoelectronic Devices, School of Science, Xi'an Jiaotong University, Shaanxi 710049, China

**Abstract:** Optical imaging through scattering media is a commonly confronted with the problem of reconstruction of complex objects and optical memory effect. To solve the problem, here, we propose a novel configuration based on the combination of ptychography and shower-curtain effect, which enables the retrieval of non-sparse samples through scattering media beyond the memory effect. Furthermore, by virtue of the shower-curtain effect, the proposed imaging system is insensitive to dynamic scattering media. Results from the retrieval of hair follicle section demonstrate the effectiveness and feasibility of the proposed method. The field of view is improved to 2.64mm. This present technique will be a potential approach for imaging through deep biological tissue.

## 1. Introduction

When a light beam propagates through scattering media, such as fog, haze, biological tissues, etc., it will be scrambled into a highly disordered complex speckle pattern because of the multiple scattering inside the media. Consequently, it will fail to image an object behind the scattering media. This is a major technical challenge in many applications, including astronomy, laser diagnosis, biomedical imaging, etc. To solve the problem, some techniques have been proposed such as adaptive optics [1] and speckle interferometry [2]. Approaches of delivering a controllable light beam behind scattering media have also been explored. In case of weak scattering, optical gate techniques [3-5] for separating ballistic (i.e., non-scattered) photons are commonly used. However, these techniques will fail when the scattering is strong. To achieve the imaging through strong scattering, some new techniques have been proposed, such as iterative optimization [6-9], optical phase conjugation [10-13], transmission matrix measurement (TM) [14-17] and speckle correlation imaging [18-23]. Work as they do, these methods just reconstruct sparse objects that has binary distributions. Meanwhile, these methods are limited to static scattering media and the memory effect.

To address the above issues, in this paper, we innovatively combine the ptychography [24, 25] and shower-curtain effect [26] for the retrieval of hidden objects. Based on the system of ptychography, a 4f optical imaging configuration is introduced to image the back surface of the scattering medium onto a detector. Owing to the shower-curtain effect, phase distortion introduced by the scattering medium is removed. The detector directly records the diffraction information of the object. Therefore, the object can be reconstructed by the extended ptychographic iterative engine algorithm. The proposed method is not limited by the memory effect. Meanwhile, the configuration is free from the sparsity of objects and suitable for dynamic scattering medium. This proposed method provides a potential approach for imaging through deep biological tissue.

## 2. Methodology

The experimental setup is shown in Fig. 1(a). A continuous wavelength (CW) laser with wavelength $\lambda$ = 532 nm is used as illumination source. After expanded and collimated, the beam illuminates a pinhole with the radius of 1 mm, which is mounted on an *x/y* stage with a precision of 10 μm. To employ the shower-curtain effect, we image the surface of the scattering medium (DG10-220, Thorlabs) onto the detector (Neo 5.5 Scmos, Andor Inc., United Kingdom) with 2560 (W) × 2160 (H) pixels of pixels size 6.5 μm (W) ×6.5 μm (H) by use of a 4f optical imaging configuration. The so-called shower curtain effect [26] is that the intensity distribution in the front surface of scattering medium is equal to the back surface. When the object is close to the scattering medium, it can be resolved on the detector with noise introduced by the scattering medium (as shown in Fig. 1(b)). In the practical case, the scattering medium placed far from the object, only diffraction pattern can be recorded (as shown in Fig. 1(c)). To reconstruct the object from the diffraction distribution, the ptychography is employed. Owing to the shower curtain, the phase distortion introduced by the scattering medium is considered as experimental noise. By moving the stage along $x$- and $y$- directions respectively, a series of diffraction patterns $I_1, I_2 \ldots I_m$ are generated.

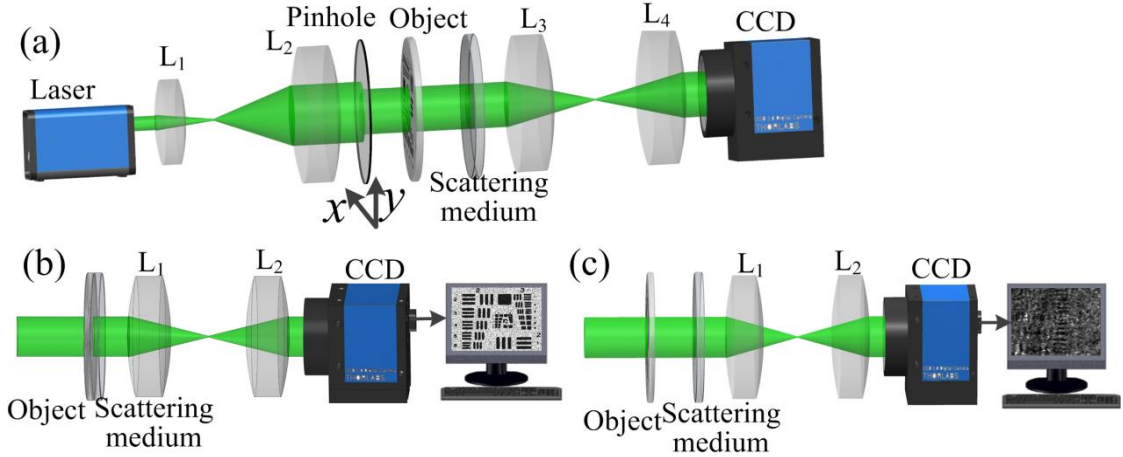

Fig. 1. The schematics of experimental setup and shower curtain effect. (a) Schematic of the experimental setup. $L_1$, $L_2$, $L_3$, $L_4$, lens; (b) Object is closely attached to the scattering medium which is imaged onto the detector; (c) Object is far from the scattering medium.

The entire process of the reconstructed algorithm is fully automated as shown in Fig. 2. Initial object $O_0(r)$ and probe wavefront $P_0(r)$ are assumed as identity matrix. After the $m^{th}$ scan, the exit wavefront at the plane of object is given by

$$E_m(r) = O_m(r) \cdot P_m(r). \tag{1}$$

By propagating the light field $E_m(r)$ to the scattering medium plane with angular spectrum (AS) theory [27], a new wavefront $E_m(r_s)$ can be generated as given by

$$E_m(r_s) = C \iint U_m(f_r) e^{-i(\frac{2\pi d}{\lambda})\sqrt{1-\lambda^2 f_x^2 - \lambda^2 f_y^2}} e^{-i2\pi(f_r r)} dr, \tag{2}$$

$$U_m(f_r) = \mathcal{F}\{E_m(r)\}. \tag{3}$$

Where $\mathcal{F}(\cdot)$ presents Fourier transform operation; $r$ and $r_s$ represent spatial coordinates of the object plane and scattering medium plane, respectively; C is a constant; $f_r$ is the spatial frequency; $d$ is the object-to-scattering medium distance.

Owing to the shower-curtain effect, the wavefront in front surface of scattering medium is addressed directly to the camera. The modulus of the wave field is next replaced with square root of the recorded speckle pattern to generate a new light field $E'_m(r_s)$:

$$E_m(r_s) = \sqrt{I(r_s)} e^{i\varphi_m(r_s)}. \tag{4}$$

An updated wavefront $E'_m(r)$ in the object plane is then calculated via an inversely angular spectrum (AS$^{-1}$) theory. By updating the current object $O_m(r)$ and probe wavefront

$P_m(r)$ using given functions, we can generate the new object $O_{m+1}(r)$ and probe $P_{m+1}(r)$. For the object wavefront, the updated function is given by

$$O_{m+1}(r) = O_m(r) + \alpha \frac{P_m^*(r)}{|P_m(r)|_{\max}^2} \cdot (E_m^{'}(r) - E_m(r)). \quad (5)$$

For the probe wavefront, it is given by

$$P_{m+1}(r) = P_m(r) + \beta \frac{O_m^*(r)}{|O_m(r)|_{\max}^2} \cdot (E_m^{'}(r) - E_m(r)), \quad (6)$$

where the parameters $\alpha$ and $\beta$ are usually set as 1 [28].

The process continues until all recorded patterns are employed to update the object and probe wavefronts. After that, the whole process is considered as a single iteration. To quantify the accuracy and convergence of the algorithm, the sum-squared error (SSE), defined as

$$SSE = \frac{\sum [|A_m(r_s)| - \sqrt{(I_m(r_s))}]^2}{\sum I_m(r_s)} \quad (7)$$

is employed after each iteration, where the function $A_m(r_s)$ indicates the calculated amplitude distribution, and the distribution $I_m(r_s)$ represents recorded intensity. The criterion value of $\varepsilon$ is preset. If the criterion $SSE < \varepsilon$ is satisfied, the algorithm ends and outputs the reconstructed object $O_{m+1}(r)$ and probe $P_{m+1}(r)$. Otherwise, it continues next iteration.

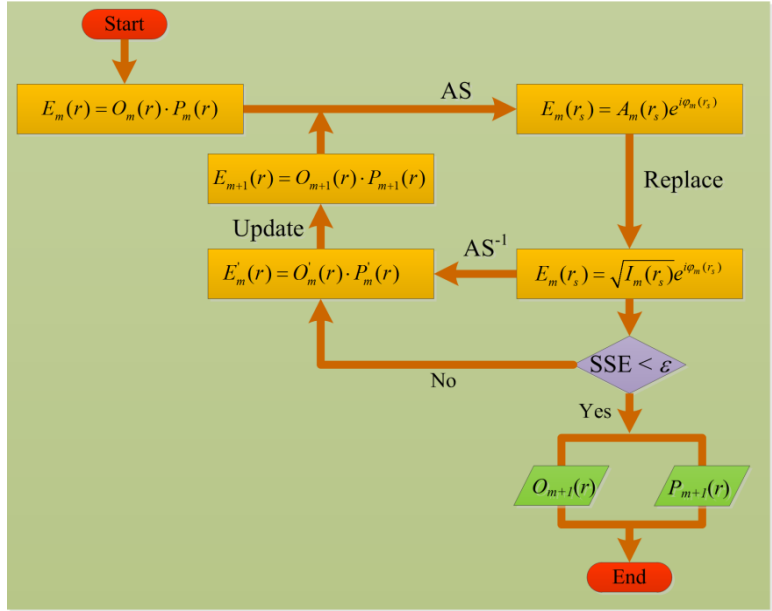

Fig. 2. The whole procedure of the reconstructed algorithm. AS represents angular spectrum; AS$^{-1}$ represents inversely angular spectrum; SSE represents the sum-square error between calculated amplitude and square root of recorded intensity at the camera's plane; $m$ represents the number of obtained patterns; $\varepsilon$ is the preset criterion value.

## 3. Experimental results and discussion

To verify the effectiveness and feasibility of the proposed method, some experiments have been carried out. The experimental setup has been shown in Fig. 1. In our experiment, by moving the stage along $x$- and $y$- directions, respectively, total 25 patterns are recorded from a grid of 5 × 5 pinhole positions with step length of $s = 300$ μm (Fig. 3(c)). The radius of the pinhole is $r = 1$ mm. The object is located at the distance of 50 mm from the scattering medium. After each speckle pattern measurement, the pinhole is moved relative to the object such that overlapping region of the object is illuminated in each position. Measurements continue until the area of interest has been covered, as shown in Fig. 3(b). For good results, an overlap rate of around 75–85% is required [24, 25]. The redundancy thereby introduced into the collected data allows retrieval of the complex amplitude field. According to geometrical optics, the overlap area can be calculated by

$$S = \frac{\arccos(\frac{s}{2r})}{90} \cdot \pi r^2 - s\sqrt{r^2 - s^2/4} \cdot \qquad (8)$$

Thus, the overlap rate between adjacent illumination areas can be given by

$$R = \frac{S}{2\pi r^2}, \qquad (9)$$

which is 80% in the experiment.

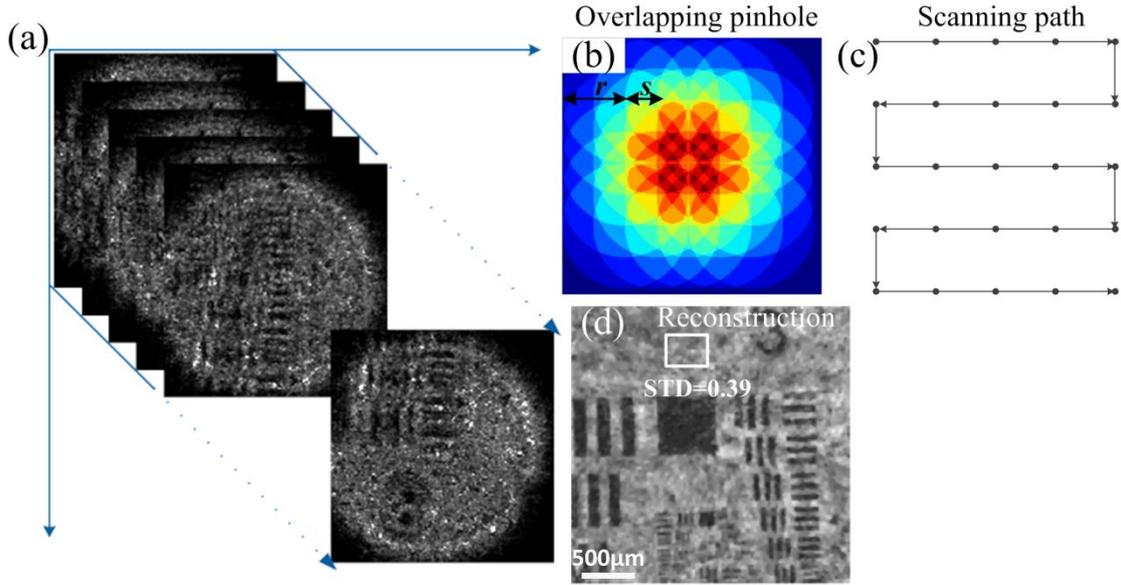

Fig. 3. The experimental result with resolution target. (a) The recorded patterns in the experiment; (b) The overlapping pinhole; (c) The scan path of pinhole; (d) Reconstructed object; $r$ and $s$ represent the radius of pinhole and the step size, respectively; STD: standard deviation.

A series of recorded patterns are shown in Fig. 3(a). To reduce execution time, only non-zero useful information with 508 × 508 pixels of each pattern is employed in the reconstruction process. In the iteration, the criterion value $\varepsilon$ is set to be 0.01. After 50 iterations, the reconstructed object is shown in Fig. 3(d). In speckle correlation imaging, the field of view (FOV) is restricted by the memory effect range of the scattering medium, resulting in a small FOV of 0.9 mm [19]. In proposed method, the FOV is 2.64 mm, which is experimentally extended beyond the memory effect. Meanwhile, the FOV can be further enlarged by scanning the pinhole extensively to record more patterns.

To exploit the performances of the proposed method for imaging the biological sample, the object is replaced with hair follicle section (Fig. 4(b)) under the same experiment. Figure 4(a) presents sub-images with the size of 508 ×508 pixels. The reconstructed object is shown in Figs. 4(c). By comparing the object (Fig. 4(b)) and the reconstruction (Fig. 4(c)), these structures which are marked by dash rectangles are almost the same. But it will fail to reconstruct those with speckle correlation technology because of the limitation of the sparsity of objects [19, 22, 23]. Therefore, the proposed method further overcomes the requirement of sparse objects and provides a potential approach for biological imaging through scattering medium.

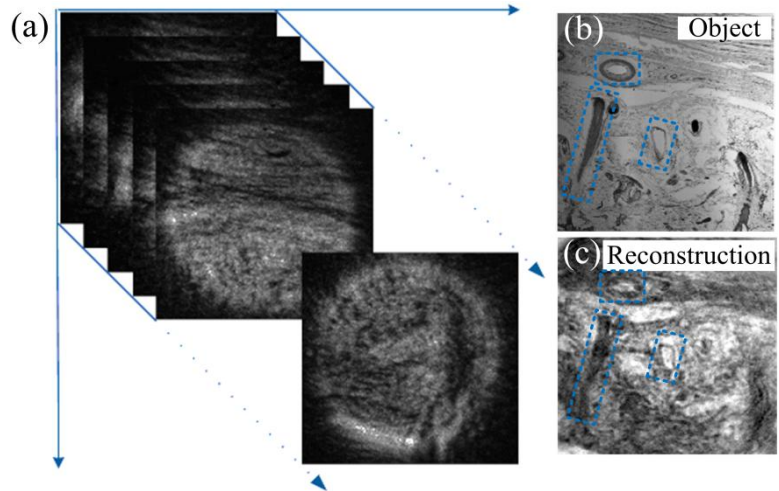

Fig. 4. The experimental result with hair follicle section. (a) The parts of recorded patterns in the experiment; (b) Object imaged by conventional microscope; (c) Reconstructed biological sample.

Another key feature of the shower-curtain phenomenon is that any phase aberration introduced by the scattering medium does not influence the intensity distribution of the detector. Thanks to this phenomenon, in our imaging system, the scattering medium can be effectively viewed as a screen where images are projected and recorded. Thus, the scattering medium is not strictly required to be static during the process. To effectively verify this performance, we still employ the same scattering medium as previous experiment and rapidly rotate it (50revs/s) using an electric motor to simulate the dynamic scattering medium. Meanwhile, total 25 speckle patterns in accordance with static way are recorded. As expected, we could accurately reconstruct the objects, as shown in Figs. 5(b) and (c). Fig 5(a) is the scattering medium. Interestingly, the quality of retrieval images through rapidly dynamic scattering medium is higher than through static medium. In dynamic case, the reconstructed images carry higher single-to-noise rate. To compare noise level in static and dynamic cases quantitatively, the standard deviation (STD) of the given area, marked with white rectangles (Figs. 3(d) and 5(b)), is calculated. In static scattering medium, the background noise is calculated to be 0.39, while it is 0.22 in the dynamic situation. The results demonstrate that the noise level is decreased by a magnitude of 17%. The reason is that speckle noises are reduced by rotating scattering medium. With the speckle field intensity being averaged out, the quality of the reconstructed image is improved.

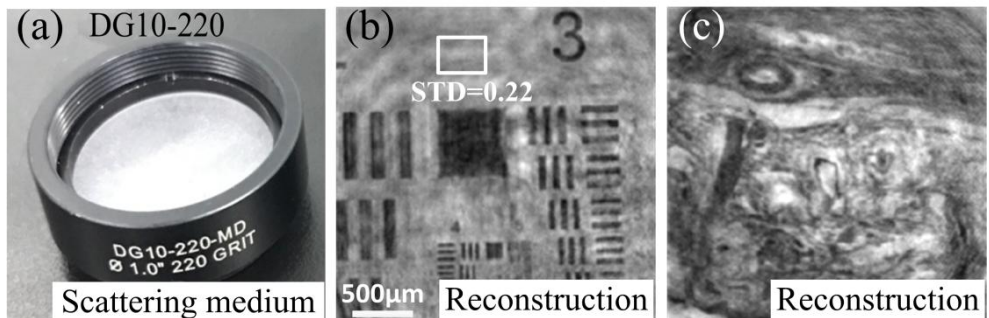

Fig. 5. The reconstruction of various samples with the dynamic scattering medium. (a) The scattering medium. (b) and (c) The reconstructions of resolution target and hair follicle.

To demonstrate the feasibility of the proposed method for imaging through biological tissue, the present scattering medium is replaced with onion tissue to repeat above experiment. The hidden objects are similarly retrieved with high fidelity by this method (see Figs. 6(b) and (c)). Thus, the objects with various shapes and complexities through dynamic scattering medium can be reconstructed robustly. Therefore, the dynamic property of practical scattering media can be exploited, rather than fought against, to image hidden objects.

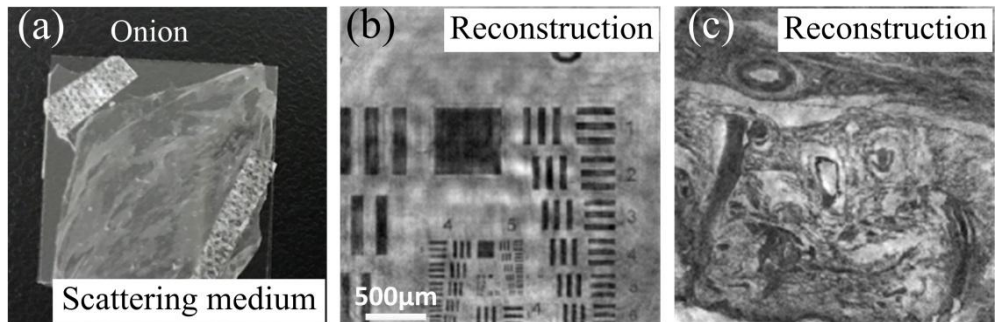

Fig. 6. Experimental imaging through dynamic onion tissue. (a) The onion tissue is employed as scattering medium. (b) and (c) The reconstructions of resolution target and hair follicle.

## 4. Conclusion

In summary, we have proposed an imaging configuration through scattering media with a large field of view. By combination of ptychography and shower-curtain phenomenon, the non-sparse object can be reconstructed with the help of a series of patterns recorded from multiple overlapping illuminations on the sample. The redundancy introduced into data can be exploited during the reconstruction. In our experiment, the field of view (FOV) is extended beyond the range of the memory effect and it can be further enlarged by scanning the pinhole extensively to record much more patterns. In addition, the utilization of shower-curtain effect makes the method insensitive to dynamic turbid media. Thus, the dynamics of practical scattering media can thus be exploited, rather than fought against, to image hidden objects. Furthermore, the samples, used in our experiment, are significantly large and realistic. These advantages provide a potential approach for practical biomedical application.

**Funding**

This research is supported by Opening Foundation of Key Laboratory of Laser & Infrared System (Shandong University), Ministry of Education and the Natural Science Foundation of China (NSFC) under grant Nos. 61705256.